# New Model System for a One-Dimensional Electron Liquid: Self-Organized Atomic Gold Chains on Ge(001)

J. Schäfer, C. Blumenstein, S. Meyer, M. Wisniewski, and R. Claessen

*Physikalisches Institut, Universität Würzburg, 97074 Würzburg, Germany*

Unique electronic properties of self-organized Au atom chains on Ge(001) in novel c(8×2) long-range order are revealed by scanning tunneling microscopy. Along the nanowires an exceptionally narrow conduction path exists which is virtually decoupled from the substrate. It is laterally confined to the ultimate limit of single atom dimension, and is strictly separated from its neighbors, as not previously reported. The resulting tunneling conductivity shows a dramatic inhomogeneity of two orders of magnitude. The atom chains thus represent an outstandingly close approach to a one-dimensional electron liquid.

The physics of electrons in low dimensions is vastly different from the conventional Fermi liquid picture, and condensed matter theory predicts exotic many-body scenarios. Most prominent for one-dimensional (1D) systems is the emergence of the Luttinger liquid which results from a decoupling of the spin and charge degrees of freedom [1]. Requirements for experimental realizations pertain to the fragility of the 1D regime, which can be affected by phonons, and may eventually be destroyed by coupling to the other dimensions [2]. Indication for spin-charge separation has been found in carbon nanotubes [3], lithographic semiconductor channels [4] and linearly bonded crystals [5]. A complication is the occurrence of a charge density wave (CDW), which can significantly alter the electronic states [6]. Moreover, wave function overlap with neighboring conduction paths as in crystals, or into supporting layers of synthetic structures is a significant deviation from the ideal 1D case. The quest for well-defined 1D systems is thus ongoing.

Promising candidates in this respect are atom chains. Gold break-junctions exhibit a suspended line of a few atoms [7]. On solid surfaces, metal atom chains can be assembled artificially with a tunneling tip up to ~20 atoms length [8]. Such short segments are governed by quantum well states. More preferable are self-organized chains on semiconductor surfaces with potentially infinite extent. Yet only a limited set of 1D reconstructions is known. Ground-breaking examples are In atoms on Si(111) [9] and Au atoms on Si(557) [10]. These chains are, however, structurally not sufficiently separated. On high-index Si, the Au atoms are even submerged into the terraces [11]. It has been speculated that Au/Si(557) exhibits spin-charge separation [10]. In subsequent work this claim has not been substantiated [12,13], and simple band structure was found to explain the observations.

Searching for better 1D character, the (001) surface deserves particular attention. Pt atoms induce narrow nanowires on Ge(001) which exhibit sharply localized states [14]. Density-functional theory relates this to Pt d-orbitals, which lead to a marginal tunneling conductivity at room temperature [15]. Gold atoms also grow in self-organized manner. A (4×2) phase was reported for excessive coverage of 1.5 monolayers (ML), while potential metallicity remained unknown [16].

In this Letter, we report on Au chains assembled on Ge(001) by low-coverage growth in novel c(8×2) long-range order. Analyzed by scanning tunneling microscopy (STM), metallic charge is spread out in 1D direction, unaffected by substrate bonds. The charge on the nanowire resides within single atom dimensions. This correlates with a dramatic lateral decline of the tunneling conductivity. Photoemission identifies a corresponding Au-induced metallic band. This renders the c(8×2) Au chains a 1D model system with unexcelled confinement.

Experimentally, n-doped Ge(001) substrates were chemically etched and flashed in ultra-high vacuum. Gold was evaporated onto the sample kept at 500 °C, thereby favoring ordered c(8×2) nanowire growth. The optimum Au coverage determined with a quartz crystal monitor was slightly above 0.5 ML. STM was performed at room temperature with an Omicron instrument.

Self-organized Au nanowires cover the whole substrate area, see Fig. 1(a). The reconstruction runs up to occasional terrace edges of the Ge(001) wafer, which has an alternating stacking sequence rotated 90°. Thus there is no principal length limitation to the nanowires, which easily extend from several 1000 Å into the μm-range. They are aligned in parallel with very even charge density in 1D direction. Obviously the dimer rows found on the (001) surface guide the chain formation during self-organization. The lateral chain spacing amounts to 16 Å, corresponding to 4× the Ge atom distance of the unreconstructed surface (a = 4.0 Å). As key feature, the linear ridges are well separated, with a pronounced minimum in-between. In close-up measurements as in Fig. 1(b), individual atoms are not identified for most tunneling conditions. This must be attributed to a strong 1D delocalization of the electron states.

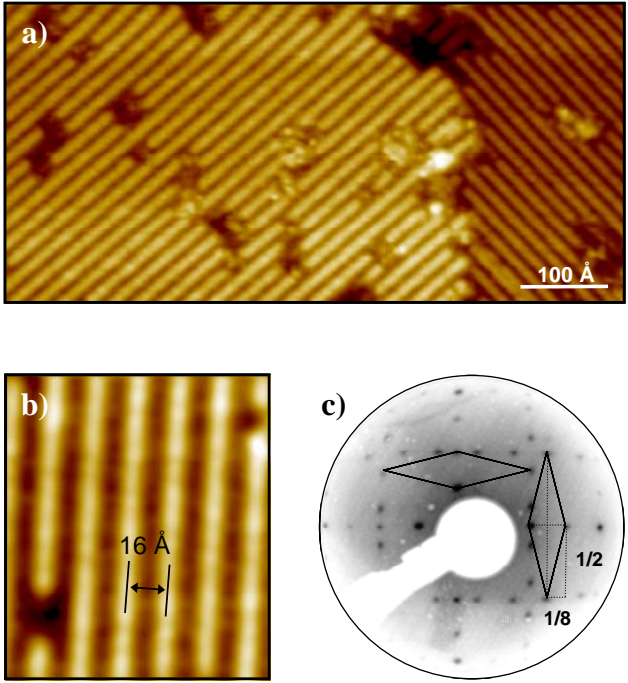

FIG. 1 (color online). STM images of Au/Ge(001). a) Overview of nanowires (720 Å × 350 Å, +1.8 V). b) Close-up (100 Å × 120 Å, +1.6 V). The nanowire spacing is 16 Å, with even intensity in 1D direction. c) LEED image at 29 eV, showing a c(8×2) structure, with dual domains.

The existence of long-range order is readily accessed by low-energy electron diffraction (LEED). The LEED image Fig. 1(c) shows a well-defined c(8×2) reconstruction, including the two surface domains. This is in contrast to Wang et al. [16] who reported a (4×2) LEED pattern. The abundant Au coverage of 1.5 ML and a low substrate temperature must be held responsible for that phase. The authors found a lot of defects in the STM data, with merely incidental occurrence of c(8×2) units. Such growth regime has been carefully avoided.

Inspection of STM images for a unit cell as in Fig. 2(a) with bias +1.5 V reveals a weak vertical undulation on the scale of 0.3 Å. It reflects the lattice periodicity of $2a = 8$ Å along the chain. Neighboring chains are shifted by 180° which accounts for the centered c(8×2) unit cell. The difficulty to observe this periodicity in STM conceivably has its reason in the delocalized charge cloud which blurs the contrast from the deeper embankment.

For the lateral *electron confinement*, STM yields an astoundingly sharp top contour of the nanowire, as in Fig. 2(b) with a rather large height of ~ 1.3 Å. In order to deduce the effective lateral dimension of the conducting filament, the resolution broadening by the STM tip needs to be considered. It originates from the tip diameter and the tunneling gap on either side of the nanowire. In the attempt to study a single Au atom of 2.9 Å diameter, one has thus to consider the tip atom as well as the tunneling gap of similar magnitude. Exact calculations yield a Gaussian-like profile, with 8.9 Å full width half maximum (FWHM) for a Na atom on a metal surface [17]. The apparent width of an atom scales slightly with the density of states [18].

Detailed profile analysis in Fig. 2(c) yields 7.2 ± 0.3 Å FWHM for the Au nanowires. In comparing to pertinent data, a curve for a *single Au atom* on an insulating alumina layer [19] is overlaid as curve (B) with ~9.0 Å FWHM. The calculated behavior of a Na atom [17] yields curve (C) with 8.9 Å FWHM. It virtually coincides with that for the single Au atom. Our measured curve of the Au nanowires is slightly smaller, and even within experimental error certainly not larger than the single atom profiles. To contrast this with an Au *dimer*, we have overlaid curve (D) recorded on an alumina substrate [19]. The deduced Au-Au spacing of 5.6 Å will vary with substrate, and may approach the bulk Au bond length of 2.9 Å on metals [20]. In any case, the resulting width of ~16 Å FWHM for this example is far beyond the current experimental observation.

This suggests that the top charge cloud is confined laterally comparable to the size of a *single* atom. We emphasize that STM detects valence orbitals rather than atom positions, so that such distribution can also reflect the spill-out from deeper layers or a bond between two atoms. Of relevance to the 1D electron liquid is solely

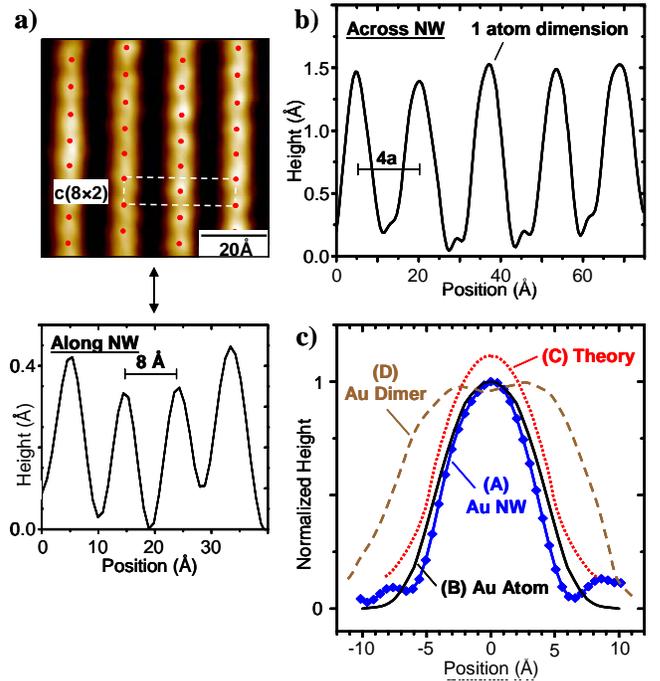

FIG. 2 (color online). a) STM image at 1.5 V, 0.5 nA with weak modulation corresponding to c(8×2) unit cell. Length profiles confirm 8 Å periodicity. b) Lateral height profile of nanowires at -2.1 V. c) Lateral profile comparison. A: nanowire data (center profile Fig. 2(b)), ~7.2 Å FWHM. B: Single Au atom on alumina [19]. C: Calculation for Na atom on metal surface [17], offset for clarity. D: Au dimer on alumina [19]. The nanowire profile matches best with the single atom data.

ent nanowire width is essentially bias-independent. We conclude that two key factors govern the images: i) the deeply modulated ridges reflect a true *topographic* component, and ii) the consistently even chain appearance requires 1D electron levels with density of states (DOS) both below and above the Fermi level $E_F$.

By contrast, in preceding STM measurements of nanowires like Pt on Ge(001) [14], variation of bias led to the observation of sideways bonds. In the present case, no such additional bonds emerge. Instead, the nanowires appear unusually homogeneous and structureless. The only viable explanation is that a rather *delocalized 1D electron system* resides on top of the nanowires, whose charge cloud dominates the tunneling signal.

Regarding the low-energy properties, information about the local DOS across $E_F$ is obtained from tunneling spectroscopy as in Fig. 4(a). It shows the differential conductivity obtained on the top ridge (averaged along the chains). A substantial tunneling conductivity exists around $E_F$ and directly proves the *metallic character*. When measured between the nanowires, the tunneling conductivity is dramatically reduced, see curve in Fig. 4(a). We have taken great care to position the tip exactly,

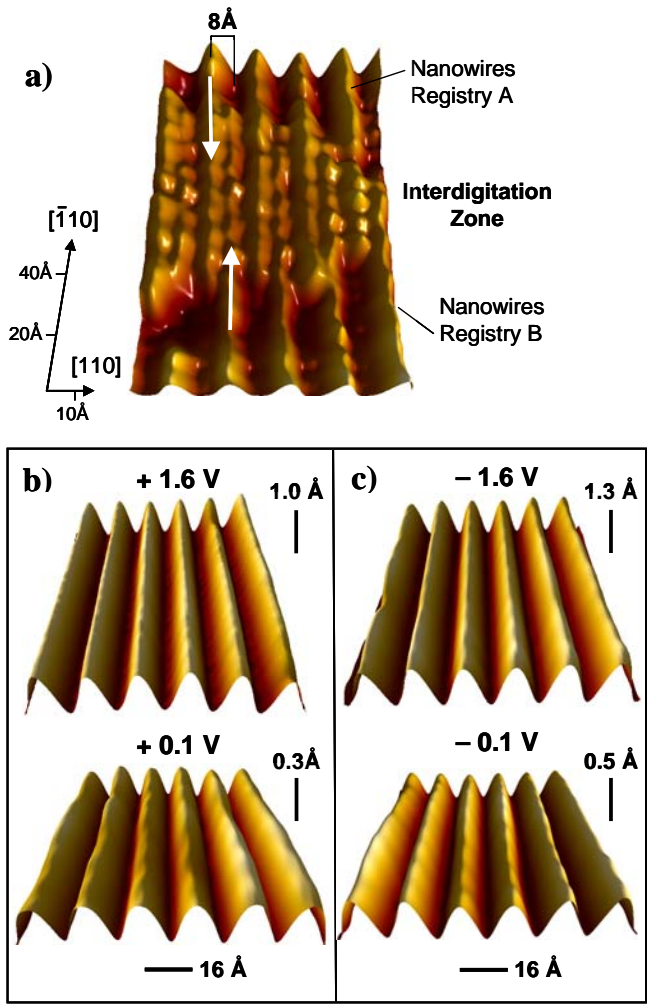

FIG. 3 (color online). a) STM image at -0.6 V, 70 Å × 100 Å of defect region with nanowire interdigitation, providing evidence for a strict spatial separation. b) Bias series of STM images (100 Å × 100 Å), with positive bias, c) negative bias. It shows persistence of identical structure independent of bias with sharp ridge and 1D charge delocalization.

the extreme spatial confinement.

Concerning the spatial separation of the chains, one may take advantage of a particular type of defective growth, namely an interdigitation of nanowires as in Fig. 3(a). This occurs at exposed terrace edges at slightly increased coverage. In addition to the regular spacing of 16 Å, the central part shows chains *interdigitated* with only half the lateral spacing. Any nanowire embankment relating to reconstructed Ge substrate bonds must thus be clearly less than 8 Å wide. It demonstrates that the atom chains are rigorously separated from each other.

In scrutinizing the charge spread in 1D direction, a sharp nanowire contour is consistently obtained when the bias is varied from positive to negative, as in Fig. 3(b) and (c). The charge density reaches a peak along the nanowire, falling off to both sides. The shape at 0.1 V bias is independent of polarity, and remains essentially unchanged for images at 1.6 V (either sign). The appar-

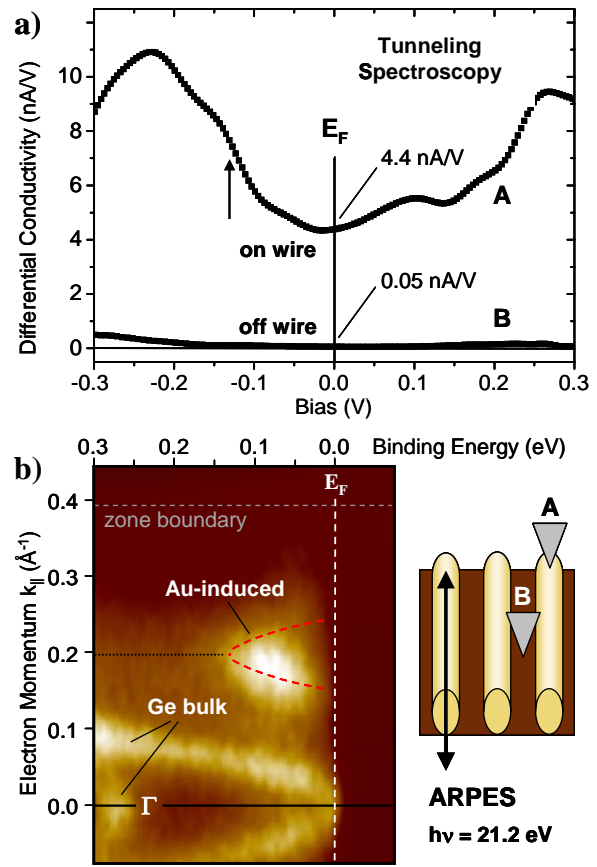

FIG. 4 (color online). a) Differential conductivity on nanowire top and in-between at 300 K. The ratio amounts to ~ 2 orders of magnitude, reflecting a pronounced inhomogeneity. b) ARPES along 1D direction with hν = 21.2 eV (T ~ 80 K). A metallic Au-induced electron band is identified.

as not to include signal from the sloping side of the nanowires. The resulting huge inhomogeneity of a factor of ~$10^2$ is a direct reflection of how well the charge is confined to the chain. The determined value may still be an underestimate due to tip broadening. We are not aware of similar data on other nanowire systems.

The existence of extended Bloch states in chain direction can be directly probed by angle-resolved photoelectron spectroscopy (ARPES). The spectra were recorded *in-situ* with He-I excitation (21.2 eV) and a Phoibos-100 analyzer. The data in Fig. 4(b) show a scan in nanowire direction through the Γ-point of the surface Brillouin zone. With a probe depth of a few monolayers, ARPES also contains information from the substrate. At the zone center, the Ge bulk band is seen as downward parabola with a maximum near $E_F$. Its spin-orbit split-off band is detected 0.3 eV below. More importantly, a shallow band is seen with a minimum of ~ 130 meV below $E_F$, challenging the experimental resolution (~50 meV). It carries very roughly the shape of an upward parabola extending through $E_F$. Such *metallic* band is not known from bulk Ge nor from the clean Ge surface [21], and must be ascribed to the Au-induced surface reconstruction. The associated DOS leads to features in the tunneling conductivity curve, where a shoulder relates to the band minimum.

For a study of spectral weight signatures, one may ask whether finite size effects play a role. For a 1000 Å chain, with an estimated free-electron Fermi velocity of roughly $6\times10^4$ m/s, the resulting level spacing is of the order of 1 meV, and further decreasing with length. The observation of Luttinger power-law behavior will thus not be obscured. Notably, as inherent advantage of this surface system, short chains limited by random defects can be intentionally selected by STM to test theories of bounded Luttinger liquids [22].

Concerning potential Luttinger liquid physics with power-law scaling around $E_F$ [1,2,22], there are first indications compatible with such scenario: i) the spectral weight in ARPES at 80 K is significantly reduced towards $E_F$ in a ~ 30 meV region, and ii) there is no evidence of long-range CDW ordering at 80 K in LEED. This is consistent with a strict 1D regime in which a CDW phase will be suppressed due to fluctuations. So one may speculate whether the system exhibits the theoretically predicted properties at low temperature. This should become visible by low-temperature tunneling spectroscopy, where non-Fermi liquid behavior leads to a recess in the differential conductivity at $E_F$ [23].

The Au chains on Ge(001) are thus distinctively different from previously described surface reconstructions which involve a lesser structural confinement. The In/Si(111) nanowires [9] consist of four metal atoms embedded between Si rows, and exhibit a CDW below ~130 K. In Au/Si(557) and Au/Si(553) [12,24], bonding to and between substrate atoms seems to govern the electronic properties [11,25]. These systems show CDW condensation already slightly below room temperature [24], which reflects significant lateral coupling. In stark contrast, the present Au chains are of single atom lateral extent. Their conduction is seemingly unaffected by the Ge(001) substrate, which is insulating at the plain surface [21]. Moreover, the conduction channels are structurally extremely well separated.

In conclusion, we find that the Au atom chains on Ge(001) embody a unique 1D electron system with unprecedented properties. On one hand, it comprises an almost fully delocalized charge distribution in chain direction, while on the other hand the confinement is laterally so narrow that the atomic limit has been reached. The system should thus provide new stimulus for studying unconventional physics in a 1D electron liquid.

The authors acknowledge discussions with F. Bechstedt and E. Tosatti, and support by the Deutsche Forschungsgemeinschaft (grant Scha1510/2-1).